\documentclass[twocolumn, showpacs, preprintnumbers, amsmath, amssymb]{revtex4}

\usepackage{graphicx}% Include figure files
\usepackage{dcolumn}% Align table columns on decimal point
\usepackage{bm}% bold math
\usepackage{axodraw}
\usepackage{epsf}

\newcommand{\be}{\begin{equation}}
\newcommand{\ee}{\end{equation}}

\newcommand{\eea}{\end{eqnarray}}
\newcommand{\bml}{\begin{mathletters} \baselineskip 10pt}
\newcommand{\eml}{\baselineskip 12pt \end{mathletters}}
\newcommand{\nn}{\nonumber}

\newcommand{\m}{{\scriptscriptstyle -}}
\newcommand{\p}{{\scriptscriptstyle +}}

\newcommand{\pprime}{\prime \prime}

\newcommand{\pad}[2]{\frac{\partial #1}{\partial #2}}

\begin{document}

\preprint{FSU--TPI/09/02}

\title{The Light--Cone Effective Potential}% Force line breaks with \\

\author{Thomas Heinzl}
%\altaffiliation[Also at ]{Physics Department, XYZ University.}
\email{heinzl@tpi.uni-jena.de}
\homepage{http://hpcs3.tpi.uni-jena.de/~sbz/}


\affiliation{Theoretisch--Physikalisches Institut,
Friedrich--Schiller--Universit\"at Jena\\
Max--Wien--Platz 1, 07743 Jena, Germany}

%\affiliation{

\date{\today}% It is always \today, today,

\begin{abstract}
It is shown how to calculate simple vacuum diagrams in light--cone quantum
field theory. As an application, I consider the one--loop effective
potential of $\phi^4$ theory. The standard result is recovered both with
and without the inclusion of zero modes having longitudinal momentum
$k^\p = 0$. 
\end{abstract}

\maketitle

\section{\label{sec:1}Introduction}

There has been a recent debate concerning the evaluation of certain
$S$--matrix elements  using light--cone quantum
field theory where canonical commutators are postulated on null--planes
tangent to the light--cone. Taniguchi \textit{et
al.}~\cite{taniguchi:01} have considered a one--loop scattering
amplitude within $\phi^4$ theory in two dimensions, given
by the \textit{finite} integral,
\be
\label{SCATTER}
  M(p^2) \equiv -i \int \frac{d^2 k}{(2\pi)^2}  \frac{1}{k^2 - m^2}
  \frac{1}{(k+p)^2 - m^2} \;.
\ee
Here, the external momentum $p$ denotes the momentum flowing through the
diagram. The authors of \cite{taniguchi:01} were particularly interested
in the case where the longitudinal momentum component, $p^\p \equiv p^0
+ p^3$, vanishes, entailing that $p^2 = p^\p p^\m = 0$. The main claim
of \cite{taniguchi:01} was that the associated scattering amplitude
\textit{vanishes}, $M (0) = 0$, if the techniques of discretized
light--cone quantization  (DLCQ) \cite{maskawa:76,pauli:85a}
are used. Its application in \cite{taniguchi:01} amounts to cutting off
$|k^\p|$ for small values, $|k^\p| > \delta$. So the finding of
\cite{taniguchi:01} actually was that $M_\delta (0) = 0$, with $
M_\delta$ denoting the integral (\ref{SCATTER}) with small--$k^\p$
cutoff.  

On the other hand,  standard covariant perturbation theory yields
(e.g.~via Wick rotation),
\begin{equation}
\label{COVRES} 
  M(0) = -i \int \frac{d^2k}{(2\pi)^2} \frac{1}{(k^2 - m^2 + i
  \varepsilon)^2} = \frac{1}{4\pi m^2} \; .
\end{equation}
Thus, there seems to be a paradox. 

The conclusions of \cite{taniguchi:01} were questioned shortly
afterwards by  the authors of \cite{harindranath:02}. They studied the
behavior of $M (p^\p p^\m)$ for $p^\p > 0$, calculated the integral in
noncovariant (Hamiltonian) light--cone perturbation theory and performed
the limit $p^\p \to 0$ at the end.  Their result was that the scattering
amplitude $M(p^2)$ \textit{does} approach the correct limit (\ref{COVRES}).

\textit{If} covariant and light--cone perturbation theory are
equivalent, then it should be possible to reproduce (\ref{COVRES}) by
evaluating the integral in LC coordinates. In this case, (\ref{COVRES})
becomes
\be
\label{LC_AMP1}
  M_{\mathrm{LC}} \equiv -i \int \frac{dk^\p \, dk^\m}{8 \pi^2}
  \frac{1}{(k^\p k^\m - m^2 + i \varepsilon)^2}  \; .
\ee
The crux is the $k^\m$-integration which is somewhat tricky.
Fortunately, this issue has been  analysed long ago by Yan in
\cite{yan:73b} who noted the following. Obviously, one wants to use
contour methods to perform the $k^-$--integration. Apparently, one can
always choose the contour such that the double pole, $k^- = m^2 /
k^+$, can be avoided. While this is true for $k^+ \ne 0$, the argument
breaks down for $k^+ = 0$. This suggests that the whole contribution
to the integral comes from the `zero mode', $k^+ = 0$, the integrand
thus being proportional to $\delta(k^+)$. This is indeed what Yan has
found,
\be
\label{YAN}
  \int dk^\m \, (k^\p k^\m - m^2 + i \varepsilon)^{-2} = \frac{2 \pi i}{m^2}
  \, \delta(k^\p) \; ,
\ee
which is most easily shown using Schwinger's parametrization to
exponentiate denominators. Plugging (\ref{YAN}) into (\ref{LC_AMP1})
correctly yields (\ref{COVRES}). 

At this point it becomes completely obvious what is going on 
in DLCQ. By cutting off the small--$k^\p$ region one is removing the
support of the delta function in (\ref{YAN}) so $M_\delta (0) = 0$ for
all $\delta > 0$. In other words, the limit $\delta \to 0$ is nonuniform.
In any case, the correct way to proceed is to either keep the external
momentum $p^\p$ finite as in \cite{harindranath:02} or to use Yan's
formula (\ref{YAN}). 

The discussion above might seem somewhat academic. However, light--cone
zero modes are of relevance in a broad range of physical problems. For
instance, they have been discussed in the matrix formulation resulting
from DLCQ of M--theory (see \cite{banks:99} and references
therein). Within phenomenological applications, zero modes show up as
end--point singularities of Feynman graphs \cite{melikhov:02} which may
lead to a breakdown of collinear factorization
\cite{braun:02}. Amplitudes with vanishing external momenta arise in
particular, whenever \textit{vacuum} properties are of concern, for
example in the phenomenon of spontaneous symmetry breaking. The elusive
character of vacuum diagrams in light--cone quantum field theory
(encoded in the phrase "the light--cone vacuum is trivial"), has never
been quite resolved. To understand spontaneous symmetry breaking in this
particular framework thus still remains a theoretical challenge (see
\cite{burkardt:97d,brodsky:97,yamawaki:98,heinzl:00} for
reviews and \cite{sugihara:98,rozowsky:00,sugihara:01} for recent discussions).

It turns out that a slight generalization of Yan's formula (\ref{YAN})
can be used to calculate a particularly relevant tool for studying
spontaneous symmetry breaking, namely the effective potential.

\section{\label{sec:2}The effective potential: ordinary coordinates}

The effective potential $V [\phi_c]$ is the leading term
of the effective action $\Gamma [\phi_c]$ in a derivative
expansion. In other words, it is the effective action evaluated for
\textit{constant} classical field $\phi_c$, divided by the volume $\Omega$ of
space--time. As $\Gamma$, on the other hand, is the generating
functional of one--particle irreducible (1PI) diagrams (proper
vertices), the effective potential can be expanded in terms of 1PI
diagrams with \textit{vanishing external momenta},
\be
\label{V_GAMMA}
  \Omega V [\phi_c] = - \sum_n \frac{1}{n!} \Gamma^{(n)} (p_1 =
  0, \ldots , p_n = 0) \phi_c^n \; .
\ee
It is this vanishing of external momenta which makes the
considerations of the introduction apply. To make this letter
self--contained we recall some basic results concerning the effective
potential. For $\phi^4$--theory in $d$ dimensions, (\ref{V_GAMMA})
becomes, in one--loop approximation, 
\be
\label{V_SUM}
   V [\phi_c] = U + i \int \frac{d^d k}{(2 \pi)^d}
  \sum_{n \ge 1} \frac{1}{2n} \left( \frac{V_0^{\pprime}}{k^2 - m^2 + 
  i \epsilon} \right)^n \; ,
\ee
where $U$ denotes the classical potential,
\be
  U [\phi_c] \equiv \frac{1}{2} m^2 \phi_c^2 + \frac{\lambda}{4!}
  \phi_c^4 \equiv \frac{1}{2} m^2 \phi_c^2 + V_0 \; ,
\ee
and the prime(s) differentiation with respect to $\phi_c$. Usually, one
evaluates (\ref{V_SUM}) by first summing and then 
integrating. The summation yields a divergent integral,
\be
\label{V_LOG}
  V [\phi_c] = U - \frac{i}{2} \int \frac{d^d
  k}{(2\pi)^d} \log \frac{U'' - k^2 - i \epsilon}{m^2 - k^2 - i
  \epsilon} \; ,
\ee
which can be evaluated most elegantly in dimensional regularisation (see
e.g.~the text \cite{peskin:95}).

In principle, for noninteger dimension $d$, all individual integrals in
the sum (\ref{V_SUM}) are regulated as well. Using Schwinger's
parametrization they can be calculated straightforwardly,
\begin{eqnarray}
\label{I_ND}
  I_{n,d} &\equiv& \int \frac{d^d k}{(2 \pi)^d}\frac{1}{(k^2 - m^2 + 
  i \epsilon)^n} \nn \\
  &=& \frac{i}{(4\pi)^{d/2}} (-1)^n \frac{\Gamma(n -
  d/2)}{\Gamma(n)} \frac{1}{(m^2)^{n - d/2}} \; .   
\end{eqnarray}
From the Gamma functions one reads off that $I_{n,d}$ is finite for
integers $n > d/2$. In particular, for even dimension $d$, there are $l
\equiv d/2$ divergent integrals. Plugging (\ref{I_ND}) into
(\ref{V_SUM}) yields the one--loop effective potential according to
\be
\label{V_INTSUM}
   V [\phi_c] = U + \frac{i}{2} \sum_{n \ge 1}
  \frac{(V_0^{\pprime})^n}{n} \, I_{n,d} \; .
\ee
Instead of evaluating this sum in full generality, we will content
ourselves with the discussion of the effective potential in $d=4$ which
is the most interesting case.

The integrals in the series (\ref{V_SUM}) are evaluated as
follows. There are two divergent ones,  
\begin{eqnarray}
  I_{1,4} &=& - \frac{i}{(4\pi)^2} \left( \Lambda^2 - m^2 \log
  \frac{\Lambda^2}{m^2} \right) \label{I_14}\; , \\
  I_{2,4} &=& \frac{i}{(4\pi)^2} \log \frac{\Lambda^2}{m^2}
  \label{I_24} \; , 
\end{eqnarray}
where, as usual, we have neglected terms that are small compared to
the covariant cutoff $\Lambda$. The finite integrals are given by the
$I_{n,4}$ from (\ref{I_ND}) with $n \ge 3$. Using the summation formula
\be
  \sum_{n \ge 3} \frac{\Gamma(n-2)}{\Gamma(n+1)}x^n = - \frac{1}{2}
  \left[ (1-x)^2 \log (1-x) + x - \frac{3}{4}x^2 \right] \; , 
\ee
the sum of the integrals (\ref{I_14}), (\ref{I_24}) and the $I_{n,4}$,
$n \ge 3$, yields the regularized one--loop effective potential 
\be
\label{V_EFF4}
  V [\phi_c] = U + \frac{1}{4} \left( \frac{U''}{4 \pi}
  \right)^2  \left( 2 \frac{\Lambda^2}{U''} +  \log
  \frac{U''}{\Lambda^2} - \frac{1}{2} \right) \; .
\ee
Exactly the same result is obtained by directly evaluating (\ref{V_LOG})
with a covariant cutoff.

To renormalize the potential, two counterterms are required corresponding to
mass and coupling constant renormalization, respectively,
\be
  V_R \equiv  V + \frac{A}{2} \phi_c^2 + \frac{B}{4!}
  \phi_c^4 \; .
\ee
The counterterms are fixed by the renormalization conditions
\begin{eqnarray}
  V_R^{\pprime} \Big|_{\phi_c = 0} &=& m^2 \; , \label{RENORM_COND1} \\
  V_R^{\pprime\pprime} \Big|_{\phi_c = 0} &=& \lambda \;
  , \label{RENORM_COND2} 
\end{eqnarray}
which finally lead to the renormalized effective potential
\be
\label{VR_D4}
  V_R [\phi_c] = U + \frac{1}{4} \left( \frac{U''}{4 \pi} \right)^2
  \left( 2 \frac{m^2}{U''} + \log \frac{U''}{m^2} - \frac{3}{2}
  \right) \; .
\ee
In what follows we will try to reproduce (\ref{V_EFF4}) and hence
(\ref{VR_D4}) using light--cone methods.

\section{\label{sec:3}The effective potential: LC coordinates}

Within the LC formalism, the calculation of the effective potential is 
a tricky business. Some early attempts
\cite{benesh:91,harindranath:88c,heinzl:92c,pinsky:94}  did not
reproduce the canonical effective potential (\ref{VR_D4}). In
\cite{xu:95}, an approach based on a \textit{bona fide} path integral has been 
adopted, which, however, did not properly include the LC constraints
into the measure. In addition, the intricacies of regulating LC diagrams 
have not been addressed. The paper that comes closests to ours
in spirit is \cite{convery:95}. These authors were discussing the
integrals $I_{n,4}$ and realized that the finite ones are entirely
dominated by the zero mode. Being not aware of Yan's formula they
suggested to define LC vacuum diagrams by letting some small amount
$p$ of external momentum flow through the diagrams and take the
limit $p \to 0$ after integration (cf.~the discussion in
the introduction).  

%\subsection{Including the zero modes}

In order to calculate the effective potential using the LC approach,
two basic difficulties have to be overcome. First, one has to find a
way to calculate the finite diagrams given by the integrals $I_n$,
$n > d/2 = l$, and second, one needs a reasonable regularization for the
divergent integrals. The first problem is solved in two
steps. We specialize to even dimensions, $d=2l$, and
calculate the integral over transverse momenta,   
\begin{eqnarray}
\label{J_NL}
  J_{n,2l} &\equiv&  \int \frac{d^{d-2}k_\perp}{(2 \pi)^{d-2}}
  \frac{1}{(k^\p k^\m - k_\perp^2 - m^2 + i \epsilon)^n} \nn \\ 
  &=& \left(-\frac{1}{4\pi} \right)^{l-1} \!\!
  \frac{\Gamma(n-l+1)/\Gamma(n)}{(k^\p k^\m  - m^2 + i
  \epsilon)^{n-l+1}} \, .
\end{eqnarray}
This formula is only relevant in more than two dimensions ($l>1$) and
yields a finite answer for integers $n > l-1$. The integrals defining the
one--loop effective potential are then
\begin{equation}
  I_{n,2l} = \frac{1}{8 \pi^2} \int dk^\p dk^\m
  J_{n,2l}  \;. \label{INT_J}
\end{equation} 
The $k^\m$--integration is done using a slight generalization of Yan's
formula (\ref{YAN}), 
\be
\label{YAN_2L}
  \int dk^\m (k^\p k^\m - m^2 + i \epsilon)^{-p} = 2 \pi i \, 
  \delta(k^\p) \frac{(-1)^p}{p-1} \frac{1}{(m^2)^{p-1}} \; ,
\ee
which holds for $p>1$ and can be proved by differentiating (\ref{YAN})
$p-2$ times with respect to $m^2$. The remaining $k^\p$--integration in
(\ref{INT_J}) immediately leads to (\ref{I_ND}). All finite integrals
are reproduced in this way. Thus, all that remains to be discussed are
the divergent integrals which, in $d=4$, are  $I_{1,4}$ and $I_{2,4}$. The
latter is simpler,  as (\ref{J_NL}) applies ($l=n=2$) and results in  
\be
\label{I_24_LC}
  I_{2,4} = \int \frac{d^4 k}{(2\pi)^4} \frac{1}{(k^2 - m^2 +
  i\epsilon)^2}  = - \frac{1}{4\pi} I_{1,2} \; .
\ee
The integral $I_{1,2}$ is the light--cone tadpole in $d=2$,
\be
\label{LC_TADPOLE_2}
  I_{1,2} = \frac{1}{8 \pi^2} \int_{(\Lambda)} dk^\p dk^\m
  \frac{1}{k^\p k^\m - m^2 + i \epsilon} \; ,
\ee
where by $(\Lambda)$ we have indicated a regularization prescription
(cutting off small and large values of $k^\p$) which will be specified in a
moment. In any case, with $|k^\p| > \delta$, the $k^\m$--integration
\textit{can} be done by closing the contour, so that 
(\ref{LC_TADPOLE_2}) becomes 
\be
  I_{1,2} = - \frac{i}{4 \pi} \int_{(\Lambda)} dk^\p \,
  \frac{\theta(k^\p)}{k^\p} \; 
  . 
\ee
It has been repeatedly noted by many people
\cite{harindranath:88c,dietmaier:89,griffin:92c,itakura:99} that the
cutoffs for small and large $k^\p$ have to be related via parity,
\be
\label{IMC+}
  \delta \equiv \frac{m^2}{\Lambda} \le k^\p \le \Lambda \; , 
\ee
leading to 
\be
\label{I_12_LC}  
  I_{1,2} = - \frac{i}{4 \pi} \int\limits_{m^2/\Lambda}^\Lambda
  \frac{dk^\p}{k^\p} = - \frac{i}{4 \pi} \log \frac{\Lambda^2}{m^2} \; ,
\ee
which reproduces the standard result. Note that zero modes have been
excluded and that the entire mass dependence of the integral comes from the
cutoff. Plugging (\ref{I_12_LC}) into (\ref{I_24_LC}) finally yields
\be
  I_{2,4} = \frac{i}{(4\pi)^2} \log \frac{\Lambda^2}{m^2} \; ,
\ee
which is (\ref{I_24}).  The $d=4$ tadpole  $I_{1,4}$ is
slightly more complicated as the $k_\perp$--integration also diverges so
that (\ref{J_NL}) cannot be used. The problem has been solved some
time ago in \cite{dietmaier:89}. The crucial point is to employ the
transverse--momentum cutoff, 
\be
\label{IMC}
  \Lambda^2 (x) \equiv \Lambda^2 x (1-x) + m^2 \; , \quad \mbox{where}
  \quad x \equiv k^\p / \Lambda \; .
\ee
By means of (\ref{IMC}) and the standard $k^\p$--cutoff
(\ref{IMC+}), the tadpole becomes
\be
\label{I_14_LC}
  I_{1,4} = - \frac{i}{(4\pi)^2} \left( \frac{\Lambda^2}{2} - m^2
  \log \frac{\Lambda^2}{m^2} \right) \; .
\ee
Rescaling $\Lambda \to \sqrt{2} \Lambda$ and omitting subleading terms 
this coincides with (\ref{I_14}). Again, zero modes have not been
invoked. The regularized and renormalized effective potentials,
(\ref{V_EFF4}) and (\ref{VR_D4}), respectively, follow. 

%\subsection{Omitting the zero modes.}

We can ask ourselves what would happen if (in the spirit of
\cite{taniguchi:01}) the zero mode contributions
were strictly cut off in all integrals be they finite or
infinite. In this case, all \textit{finite} integrals $I_n$ are zero,
as the support of Yan's delta function in (\ref{YAN_2L}) is located
outside the integration region. The regularized effective potential
in $d=4$ becomes
\be
  V_{\mathrm{nZM}} = U + \frac{1}{4} \left(\frac{U''}{4 \pi}
  \right)^2 \left(  2 \frac{\Lambda^2}{U''} + \log
  \frac{m^2}{\Lambda^2} \right)  \; , \label{V_NZM4}
\ee
where `nZM' stands for 'no zero modes'. Note that there is no
$\phi_c$--dependence in the log as this resummation effect must be 
absent. Accordingly, the result (\ref{V_NZM4}) is \textit{polynomial} in
$\phi_c$. This has peculiar consequences. Imposing the renormalization
conditions (\ref{RENORM_COND1}) and (\ref{RENORM_COND2}) we find that the
renormalized effective potential coincides with the tree level expression,
$V_{\mathrm{nZM, R}} = U$. In other words, the (renormalized) one--loop
contributions vanish, a result that is clearly false. In some sense,
this can be viewed as an unwanted consequence of assuming (or rather
enforcing) a trivial light--cone vacuum. 

There is, however, a way around this obstacle. Note that the mass
dependence appearing in the log of (\ref{V_NZM4}) is entirely due to the
cutoff. If we change this cutoff in an ad--hoc manner by replacing
\be
\label{REPLACE_M}
  m^2 \to U'' = m^2 + \frac{\lambda}{2} \phi_c^2 = m^2 + V_0'' \; , 
\ee
we obtain instead of (\ref{V_NZM4}) the regularized potential
\be
  V = U + \frac{1}{4} \left( \frac{U''}{4 \pi}
  \right)^2  \left( 2 \frac{\Lambda^2}{U''} +  \log
  \frac{m^2}{\Lambda^2} + \log \frac{V_0''}{\Lambda^2} \right) \; ,
\ee
which coincides with (\ref{V_EFF4}) up to a finite renormalization. 
The first two terms in brackets are just the contribution (\ref{V_NZM4})
where the zero modes had been discarded. The latter exclusively
constitute the  last log term. We thus can state a peculiar
finding: \textit{the resummation of the zero mode contributions results 
in a modification of the cutoff} given by the replacement
(\ref{REPLACE_M}). This replacement seems to be very much ad hoc. Its
justification will be explained in what follows.

It would have been desireable to directly calculate the resummed
expression (\ref{V_LOG}) for the effective potential using LC
coordinates. This is not straightforward. Being the log--determinant
of the fluctuation operator $\Box + U''$, (\ref{V_LOG}) diverges and
has to be regulated. We know how to do so for tadpole type Feynman
diagrams, but not for the log--determinant. 

However, there is a simple trick to relate the latter to a
tadpole. Taking the derivative with respect to $\phi_c$,
\begin{eqnarray}
  \pad{}{\phi_c} V &=& U' + \frac{i}{2} U''' \int
  \frac{d^d k}{(2\pi)^d} \frac{1}{k^2 - U'' + i \epsilon} \nn \\ 
  &=&  U' + \frac{i}{2} U''' I_{1,d} (U'') \; ,
\label{TADPOLE}
\end{eqnarray}
precisely yields a tadpole with the mass $m^2$ replaced by $U''$! For
$d=4$, the relevant integral has been calculated in (\ref{I_14_LC}). All
that is left is to plug this into (\ref{TADPOLE}) and undo the
differentiation by integrating. This precisely reproduces
(\ref{V_EFF4}).

%\section{\label{sec:4}Discussion}  

We have seen that the effective potential can be calculated both with
and without the inclusion of light--cone zero modes. Omitting the
latter, however, requires a sophisticated modification of the
regularization prescription which need not work for arbitrary vacuum
graphs (see also \cite{burkardt:93b} and \cite{wilson:94} for a similar
philosophy). In order to correctly reproduce \textit{finite} vacuum 
amplitudes, the inclusion of zero modes seems indispensable. It would be
of interest to extend the techniques presented to the light--cone
Hamiltonian formulation. Work in this direction is under way. 

%\begin{acknowledgments}

I am grateful to C.~Ford and T.~Mohaupt for illuminating discussions. I 
thank S.\ Mukhanov for the hospitality extended to me and A.\ Wipf for
continuous support. This work was funded in part by DFG under
contract Wi 777/5--1.   

%\end{acknowledgments}

%\appendix
%\section{Appendixes}

%\bibliographystyle{../../bibfiles/h-physrev}
%\bibliography{../../bibfiles/LC}

\end{document}